\documentclass[11pt]{article}

\usepackage[utf8]{inputenc}
\usepackage[english]{babel}
\usepackage{amsmath, amssymb, amsthm} 
\usepackage{graphicx}                 
\usepackage{hyperref}                 
\usepackage{geometry}                 
\usepackage[authoryear,round]{natbib}  
\usepackage{booktabs}                 
\usepackage{caption}                  
\usepackage{floatrow} 
\usepackage{tablefootnote}
\usepackage{float}
\usepackage{orcidlink}
 
\title{\textbf{A Genetic Algorithm Framework for Optimizing Three-Impulse Orbital Transfers with Poliastro Simulation}}

\author{
  Phuc Hao Do$^{1}$ \\
  \texttt{do.hf@sut.ru,\orcidlink{0000-0003-0645-0021}} \\
  \and
  Tran Duc Le$^{2}$ \\
  \texttt{let@uwstout.edu,\orcidlink{0000-0003-3735-0314}} \\
}

\date{
  $^{1}$Department of Telecommunication Engineering, Bonch-Bruevich Saint Petersburg State University of Telecommunications \\
  $^{2}$Mathematics, Statistics and Computer Science, University of Wisconsin–Stout \\[2ex]
}

\begin{document}

\maketitle

\begin{abstract}
    Orbital maneuver planning is a critical aspect of mission design, aimed at minimizing propellant consumption, which is directly correlated with the total velocity change ($\Delta V$). While analytical solutions like the Hohmann and Bi-elliptic transfers offer optimal strategies for specific cases, they lack the flexibility for more general optimization problems. This paper presents a computational framework that couples a Genetic Algorithm (GA) with the Poliastro orbital mechanics library to autonomously discover fuel-optimal, three-impulse transfer trajectories between coplanar circular orbits. We validate this framework across two distinct scenarios: a low-energy transfer from Low Earth Orbit (LEO) to a Geostationary Orbit (GEO), and a high-energy transfer to a distant orbit with a radius 20 times that of LEO. Our results demonstrate the framework's remarkable adaptability. For the LEO-to-GEO transfer, the GA precisely converges to the classical Hohmann transfer, achieving an identical $\Delta V$ of 3853.96 m/s and validating the method's accuracy. Conversely, for the high-energy transfer, the GA identifies a superior Bi-elliptic trajectory that yields a significant $\Delta V$ saving of 213.47 m/s compared to the Hohmann transfer. This fuel efficiency, however, necessitates a trade-off, extending the mission duration from approximately 1 day to over 140 years. This work demonstrates an accessible and powerful toolchain for the rapid prototyping of optimal trajectories, showcasing how combining evolutionary algorithms with open-source libraries provides a robust method for solving complex astrodynamics problems and quantifying their critical design trade-offs.
\end{abstract}

\section{Introduction}

The design of fuel-optimal trajectories is a cornerstone of modern astrodynamics, profoundly influencing the feasibility, cost, and scientific return of space missions \cite{Caleb2025A}. Propellant mass often constitutes a significant portion of a spacecraft's total mass at launch, making its efficient use a primary design constraint \cite{BiswalM2023}. The total required change in velocity, or delta-V ($\Delta V$), serves as a direct proxy for propellant consumption via the Tsiolkovsky rocket equation \citep{Tsiolkovsky1903}. Consequently, the minimization of $\Delta V$ for orbital transfers has remained a fundamental and intensely studied optimization problem \citep{Vallado2013}.

For the canonical problem of transferring a spacecraft between two coplanar circular orbits, analytical solutions provide foundational insights. The two-impulse Hohmann transfer, conceived in 1925, is celebrated for its efficiency in low-energy regimes \citep{Hohmann1925}. Later analysis revealed that for high-energy transfers, where the ratio of the final to initial orbit radii ($r_f/r_i$) exceeds approximately 11.94, a three-impulse Bi-elliptic transfer becomes the more fuel-efficient strategy, albeit at the cost of a significantly longer transfer time \citep{Sternfeld1934}.

While these classical maneuvers are analytically elegant, their applicability is limited to highly idealized scenarios. Real-world missions often involve non-coplanar orbits, atmospheric drag, third-body perturbations, and operational constraints that render these solutions suboptimal or invalid \cite{Jiang2022Aerodynamic}. This complexity necessitates the use of computational optimization techniques capable of navigating vast and rugged search spaces.

In recent years, the convergence of computational intelligence and accessible, high-fidelity simulation tools has opened new avenues for tackling these challenges. Evolutionary algorithms, such as Genetic Algorithms (GAs), have emerged as powerful, robust tools for global optimization, capable of finding near-optimal solutions without requiring gradient information or a good initial guess \citep{Goldberg1989}. The maturation of open-source scientific Python libraries makes this approach more feasible for a wider audience and addresses an increasing need for rapid mission prototyping, particularly for constellations or CubeSat missions \cite{Poghosyan2017CubeSat,Labreche2022Agile,Rodriguez2022poliastro,Holliday2019PyCubed}.

This paper presents a framework that synergizes a Genetic Algorithm with Poliastro \citep{Poliastro}, a modern open-source Python library for orbital mechanics, to autonomously find fuel-optimal, three-impulse transfer trajectories. Our primary contribution is not a novel algorithm, but the demonstration of a modern, accessible, and reproducible research framework. This approach contrasts with traditional numerical optimal control methods that often demand significant problem-specific formulation and initialization effort \citep{Betts2010}. By applying this framework to two distinct orbital energy regimes, we demonstrate its ability to:
\begin{itemize}
    \item Autonomously rediscover and validate classical optimal solutions (Hohmann and Bi-elliptic transfers) without prior knowledge.
    \item Quantify the critical trade-off between fuel efficiency ($\Delta V$) and mission duration.
    \item Showcase a practical toolchain for rapid prototyping and reproducible research in astrodynamics using open-source software.
\end{itemize}
The structure of this paper is as follows: Section 2 reviews related work, Section 3 details the problem formulation, Section 4 describes our methodology, Section 5 presents and discusses the experimental results, and Section 6 concludes the paper.

\section{Related Work}

The optimization of spacecraft trajectories has been a central theme in astrodynamics since the dawn of the space age. The body of work in this area can be broadly classified into three main categories: analytical methods, numerical optimal control, and heuristic/stochastic optimization.

\paragraph{Analytical and Semi-Analytical Methods.}
The field's foundations lie in analytical solutions derived for simplified problems. The Hohmann transfer represents the optimal two-impulse solution for coplanar circular orbit transfers \citep{Hohmann1925}. This was later expanded upon by the Bi-elliptic transfer, which demonstrated that three impulses could be more efficient for high-energy transfers \citep{Sternfeld1934}. Other notable contributions include the study of optimal multi-impulse transfers between orbits, famously addressed by Lawden's primer vector theory \citep{Lawden1963}, which provides necessary conditions for optimality. While these methods offer profound theoretical insight, their direct application is often limited to unperturbed, two-body dynamics.

\paragraph{Numerical Optimal Control.}
To handle more realistic and complex scenarios involving constraints and detailed dynamics (e.g., low-thrust propulsion, atmospheric drag), researchers have extensively employed numerical methods rooted in optimal control theory \citep{Betts2010}. These are typically divided into two classes. \textit{Indirect methods} solve the two-point boundary value problem that arises from the necessary conditions for optimality (e.g., Pontryagin's Minimum Principle). While they can produce highly accurate solutions, they are notoriously difficult to initialize and converge. \textit{Direct methods}, such as direct collocation, are generally more robust. They transcribe the continuous optimal control problem into a large-scale nonlinear programming (NLP) problem by discretizing the state and control variables over time, which can then be solved by established NLP solvers. Both approaches, however, often require significant problem-specific formulation and a good initial guess to guide the solver.

\paragraph{Heuristic and Stochastic Optimization.}
In response to the challenges of local minima and initialization sensitivity in classical optimization, heuristic and stochastic methods have become increasingly popular. These population-based algorithms, such as (GAs) \citep{Goldberg1989}, Particle Swarm Optimization (PSO), and Ant Colony Optimization (ACO), perform a global search of the solution space. Their gradient-free nature makes them well-suited for non-convex and discontinuous problems. GAs, in particular, have a long history of successful application in astrodynamics, from optimizing multi-impulse Earth-orbit transfers \citep{Pourtakdoust2005} to designing complex, multi-gravity-assist interplanetary trajectories \citep{Rocco2012}. More recent work has also explored hybrid approaches that combine the global search capabilities of GAs with the precision of local numerical methods \citep{Vasile2006}.

\paragraph{Our Contribution.}
Our work aligns with the long tradition of applying GAs to trajectory optimization. However, our primary contribution is not the proposal of a novel algorithm, but rather the demonstration of a modern, highly accessible, and reproducible research framework. By integrating the DEAP optimization library with the open-source Poliastro simulation engine, we present a complete, lightweight, scriptable toolchain. The value of this framework lies in its capacity for\textbf{ rapid prototyping}, allowing researchers to quickly set up and solve complex optimization problems without the steep learning curve or significant formulation effort associated with the traditional numerical methods described above. This paper serves as both a validation of the methodology - by showing its ability to rediscover known optimal solutions - and a practical guide for leveraging contemporary open-source tools for research in orbital mechanics.

\section{Problem Formulation}

The central problem addressed in this paper is the identification of a fuel-optimal, three-impulse trajectory for transferring a spacecraft between two specified coplanar, circular orbits. This task is framed as a constrained nonlinear optimization problem, where the objective is to minimize the total propulsive effort required for the transfer.

\subsection{Dynamical Model and State Representation}
We model the spacecraft's motion within the framework of the restricted two-body problem. The spacecraft is treated as a point mass, moving under the gravitational influence of a single, spherically symmetric central body (Earth), characterized by its standard gravitational parameter, $\mu$. The state of the spacecraft at any time $t$ is fully described by its position vector $\vec{r}(t)$ and velocity vector $\vec{v}(t)$ in an inertial reference frame. The equation of motion is given by:
\begin{equation}
    \frac{d^2\vec{r}}{dt^2} = -\frac{\mu}{r^3}\vec{r}
\end{equation}
where $r = \|\vec{r}\|$.

The initial orbit, $\mathcal{O}_i$, is a circular trajectory with a constant radius $r_i$, where the velocity is purely tangential with a magnitude given by:
\begin{equation}
    v_i = \sqrt{\frac{\mu}{r_i}}
\end{equation}
Similarly, the final target orbit, $\mathcal{O}_f$, is a circular trajectory with radius $r_f$ and velocity magnitude $v_f = \sqrt{\mu/r_f}$.

\subsection{Transfer Trajectory Model}
The transfer is modeled as a sequence of three instantaneous impulses, $\Delta\vec{v}_1, \Delta\vec{v}_2, \Delta\vec{v}_3$, which define a Bi-elliptic transfer. This maneuver utilizes two intermediate elliptical transfer orbits, $\mathcal{T}_1$ and $\mathcal{T}_2$.

\begin{itemize}
    \item \textbf{First Transfer Orbit ($\mathcal{T}_1$):} An initial impulse, $\Delta\vec{v}_1$, is applied at $\mathcal{O}_i$ to inject the spacecraft into $\mathcal{T}_1$. This elliptical orbit is defined by its periapsis radius, $r_{p1} = r_i$, and an apoapsis radius, $r_{a1} = r_b$, where $r_b$ is an intermediate radius.

    \item \textbf{Second Transfer Orbit ($\mathcal{T}_2$):} At the apoapsis of $\mathcal{T}_1$ (at radius $r_b$), a second impulse, $\Delta\vec{v}_2$, is applied. This maneuver alters the trajectory to place the spacecraft onto $\mathcal{T}_2$, which has a periapsis radius of $r_{p2} = r_f$ and an apoapsis radius of $r_{a2} = r_b$.

    \item \textbf{Final Orbit Injection:} Upon reaching the periapsis of $\mathcal{T}_2$ (at radius $r_f$), a final impulse, $\Delta\vec{v}_3$, is applied to circularize the orbit and match the velocity of the target orbit $\mathcal{O}_f$.
\end{itemize}

The sole decision variable in this model is the intermediate apoapsis radius, $r_b$.

\subsection{Optimization Problem Statement}
The overarching goal is to minimize the total propellant mass required for the transfer. The Tsiolkovsky rocket equation demonstrates that this is equivalent to minimizing the scalar sum of the magnitudes of all propulsive impulses. This sum is known as the total delta-V, $\Delta V_{\text{total}}$.

The optimization problem can therefore be stated formally as follows:

\textbf{Find} the optimal decision variable:
\begin{equation}
    r_b^*
\end{equation}

\textbf{To minimize} the objective function $J(r_b)$, defined as the total delta-V:
\begin{equation}
    J(r_b) = \Delta V_{\text{total}} = \sum_{k=1}^{3} \|\Delta \vec{v}_k(r_b)\|
    \label{eq:objective_function}
\end{equation}
where $\|\Delta \vec{v}_k\|$ is the magnitude of the $k$-th impulse, which is a function of the intermediate radius $r_b$. The specific expressions for these magnitudes are derived from the vis-viva equation and are detailed in the Methodology section.

\textbf{Subject to} the physical constraint for a valid Bi-elliptic transfer:
\begin{equation}
    r_b \ge r_f
    \label{eq:constraint}
\end{equation}
For a non-degenerate transfer, this inequality is strict, $r_b > r_f$. The case where $r_b = r_f$ represents a special degenerate case where the Bi-elliptic transfer collapses into a two-impulse Hohmann transfer.

This formulation effectively reduces the complex trajectory optimization problem to a single-variable, constrained optimization task. The nature of the objective function $J(r_b)$ is well-defined, but its global minimum must be identified within the feasible search space defined by the constraint. This structure makes the problem particularly amenable to heuristic search methods like Genetic Algorithms.

\section{Methodology}

Our methodology integrates a Genetic Algorithm with the Poliastro orbital mechanics library to solve the three-impulse orbital transfer optimization problem. The GA serves as a global search engine, proposing candidate trajectories, while Poliastro provides high-fidelity simulation to evaluate their feasibility and cost \citep{Poliastro}. This section outlines the orbital dynamics model, the Bi-elliptic transfer formulation, and the GA implementation, emphasizing the synergy between these components.

\subsection{Orbital Dynamics Model}

The spacecraft's motion is modeled within the restricted two-body problem, where the spacecraft is a point mass orbiting a central body (Earth) with standard gravitational parameter $\mu = 3.986 \times 10^{14} \, \text{m}^3/\text{s}^2$. The governing equation of motion is:
\begin{equation}
    \ddot{\vec{r}} + \frac{\mu}{r^3}\vec{r} = 0,
\end{equation}
where $\vec{r}$ is the position vector relative to Earth, and $r = \|\vec{r}\|$ is its magnitude. We use the Poliastro library (v0.17.0) for orbit propagation and maneuver calculations. Poliastro employs a robust numerical propagator to compute the spacecraft's state (position and velocity) over time, ensuring accurate simulation of elliptical and circular orbits. The initial orbit for all scenarios is a circular Low Earth Orbit (LEO) at an altitude of 400 km ($r_i \approx 6778 \, \text{km}$), with a velocity given by:
\begin{equation}
    v_i = \sqrt{\frac{\mu}{r_i}}.
\end{equation}

\subsection{Bi-elliptic Transfer Formulation}

The optimization problem focuses on finding a fuel-optimal three-impulse Bi-elliptic transfer between two coplanar circular orbits: the initial orbit with radius $r_i$ and the target orbit with radius $r_f$. The transfer consists of two intermediate elliptical orbits, $\mathcal{T}_1$ and $\mathcal{T}_2$, connected by three instantaneous velocity changes ($\Delta \vec{v}_1$, $\Delta \vec{v}_2$, $\Delta \vec{v}_3$). The sole decision variable is the intermediate apoapsis radius, $r_b$, which defines the geometry of the transfer. The maneuver sequence is as follows:

\begin{itemize}
    \item \textbf{Impulse 1 ($\Delta V_1$):} Applied at $r_i$ on the initial orbit, this prograde impulse injects the spacecraft into $\mathcal{T}_1$, an elliptical orbit with periapsis radius $r_{p1} = r_i$ and apoapsis radius $r_{a1} = r_b$. Using the vis-viva equation, the impulse magnitude is:
    \begin{equation}
        \Delta V_1 = \sqrt{\frac{2\mu}{r_i} - \frac{2\mu}{r_i + r_b}} - \sqrt{\frac{\mu}{r_i}}.
    \end{equation}

    \item \textbf{Impulse 2 ($\Delta V_2$):} Applied at the apoapsis of $\mathcal{T}_1$ ($r_b$), this impulse adjusts the trajectory to place the spacecraft onto $\mathcal{T}_2$, an elliptical orbit with periapsis radius $r_{p2} = r_f$ and apoapsis radius $r_{a2} = r_b$. The magnitude is:
    \begin{equation}
        \Delta V_2 = \sqrt{\frac{2\mu}{r_b} - \frac{2\mu}{r_b + r_f}} - \sqrt{\frac{2\mu}{r_b} - \frac{2\mu}{r_b + r_i}}.
    \end{equation}

    \item \textbf{Impulse 3 ($\Delta V_3$):} Applied at the periapsis of $\mathcal{T}_2$ ($r_f$), this retrograde impulse circularizes the trajectory to match the target orbit’s velocity. The magnitude is:
    \begin{equation}
        \Delta V_3 = \sqrt{\frac{\mu}{r_f}} - \sqrt{\frac{2\mu}{r_f} - \frac{2\mu}{r_b + r_f}}.
    \end{equation}
\end{itemize}

The total cost of the transfer, which we aim to minimize, is the sum of the impulse magnitudes:
\begin{equation}
    \Delta V_{\text{total}} = \Delta V_1 + \Delta V_2 + \Delta V_3.
    \label{eq:total_dv}
\end{equation}
Poliastro’s \texttt{Maneuver.bielliptic} function computes these $\Delta V$ values and the resulting orbital parameters with high numerical precision, leveraging its two-body propagator to ensure accurate trajectory simulation. This integration allows rapid evaluation of candidate trajectories proposed by the GA.

\subsection{Genetic Algorithm Implementation}

We employ a Genetic Algorithm, implemented using the DEAP library \citep{DEAP}, to search for the optimal $r_b$ that minimizes $\Delta V_{\text{total}}$. The GA’s key components are designed as follows:

\begin{itemize}
    \item \textbf{Chromosome Representation:} Each individual in the GA population is represented by a single real-valued gene, $\rho = r_b / r_f$, the ratio of the intermediate apoapsis radius to the final orbit radius. This normalization simplifies the search space and ensures scalability. To effectively find the optimal solution in different energy regimes, the search space for $\rho$ was adapted to each scenario. For the low-energy LEO-to-GEO transfer, the search range was set to $\rho \in [1.0, 40.0]$. The lower bound of 1.0 allows the GA to find the degenerate Hohmann transfer case ($r_b = r_f$), while the upper bound is sufficient for this regime. For the high-energy LEO-to-Far-Orbit scenario, where theory suggests a much larger intermediate radius is optimal \citep{Sternfeld1934}, the upper bound was relaxed significantly to $\rho \in [1.0, 1000.0]$. This flexibility is crucial for allowing the algorithm to discover the globally optimal strategy without being artificially constrained.

    \item \textbf{Fitness Function:} The fitness of an individual is the total $\Delta V$ computed using Equation \ref{eq:total_dv}, with the objective to minimize:
    \begin{equation}
        \text{Minimize} \quad f(\rho) = \Delta V_{\text{total}}(\rho).
    \end{equation}
    Poliastro’s \texttt{Maneuver.bielliptic} function evaluates $f(\rho)$ by simulating the transfer and computing the precise $\Delta V$ values. To handle the physical constraint $r_b \ge r_f$ (or $\rho \ge 1$), individuals with $\rho < 1$ are assigned a large penalty fitness value ($10^6 \, \text{m/s}$), effectively excluding them from selection.

    \item \textbf{GA Operators and Parameters:} The GA is configured with standard evolutionary operators:
    \begin{itemize}
        \item \textbf{Population Size:} 40 individuals.
        \item \textbf{Generations:} 30. Convergence plots, provided in Appendix A, confirm the algorithm reliably reaches a stable optimum within this limit for both scenarios.
        \item \textbf{Selection:} Tournament selection with a tournament size of 3.
        \item \textbf{Crossover:} Simulated Binary Crossover (SBX), implemented via \texttt{tools.cxBlend}, with a probability of 0.7.
        \item \textbf{Mutation:} Gaussian mutation, implemented via \texttt{tools.mutGaussian}, with a probability of 0.2 and a standard deviation of 0.1.
    \end{itemize}
\end{itemize}
The selection of these parameters was informed by preliminary experiments to ensure robust performance. To assess sensitivity, we tested variations in population size (from 20 to 60) and mutation probability (from 0.1 to 0.3), finding that the chosen configuration consistently achieved convergence to within 0.1\% of the optimal $\Delta V$ in both scenarios.

\subsection{Implementation Notes}

The integration of DEAP and Poliastro creates a seamless optimization pipeline. DEAP handles the evolutionary process - generating and evolving candidate $\rho$ values - while Poliastro evaluates each candidate’s fitness by simulating the Bi-elliptic transfer and computing $\Delta V_{\text{total}}$. The computational cost of each GA run is modest, averaging approximately 10 seconds on a standard laptop (Intel i7, 16 GB RAM), making the framework practical for the rapid prototyping of transfer trajectories. This methodology leverages the global search capabilities of GAs and the high-fidelity simulation of Poliastro, enabling autonomous discovery of optimal transfer strategies without requiring analytical initial guesses.

\section{Experiments and Results}

To validate our GA-based optimization framework, we conducted two distinct experiments comparing the performance of the GA-discovered three-impulse trajectory against the classical two-impulse Hohmann transfer. The scenarios were specifically chosen to test the framework's ability to adapt to different orbital energy regimes, where classical theory predicts different optimal strategies.

\subsection{Scenario 1: LEO to Geostationary Orbit (GEO)}

The first scenario involves a transfer from a 400 km altitude LEO ($r_i \approx 6778.00$ km) to a Geostationary Orbit ($r_f \approx 42164.00$ km). The ratio of the final to initial orbit radii is $r_f/r_i \approx 6.2$. This value is below the theoretical threshold of $\approx 11.94$, where the Hohmann transfer is known to be the most fuel-efficient. This scenario serves as a critical validation case for our algorithm.

\begin{figure}[!t]
    \centering
    \includegraphics[width=0.8\textwidth]{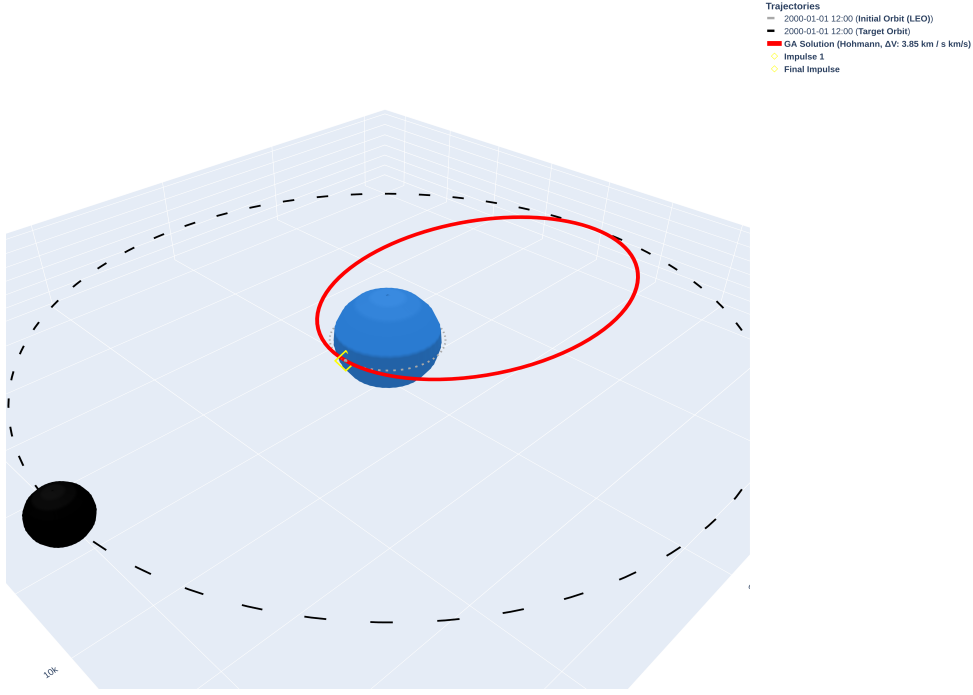}
    \caption{Optimal trajectories for the LEO-to-GEO transfer. The GA-optimized path (red) perfectly overlays the classical Hohmann transfer, demonstrating the algorithm's precise convergence to the known optimal solution.}
    \label{fig:geo_results}
\end{figure}

The Genetic Algorithm converged robustly over 30 generations, consistently locating the global optimum. The detailed results, presented in the "LEO to GEO" section of Table \ref{tab:detailed_results}, show the GA identified a trajectory with a total $\Delta V$ of 3853.96 m/s. This is identical to the cost of the classical Hohmann transfer. Analysis of the optimal individual found by the GA reveals that the intermediate apoapsis ratio ($r_b/r_f$) converged precisely to 1.00, which collapses the three-impulse maneuver into a perfect two-impulse Hohmann transfer. This result strongly validates our framework's ability to precisely identify the true analytical optimum in a well-defined problem space. Figure \ref{fig:geo_results} provides a visual confirmation of this exact convergence.

\subsection{Scenario 2: LEO to a Distant Orbit}

The second scenario was designed to test the framework in a high-energy regime where the Bi-elliptic transfer is theoretically superior. The target orbit was set to a circular orbit with a radius 20 times that of the initial LEO ($r_f/r_i = 20$), placing it significantly above the 11.94 threshold.

In this case, the GA's performance was again remarkable. It converged on a solution with a very large intermediate apoapsis radius of over 118 million km ($r_b/r_f \approx 872.00$)—a result made possible by the expanded search space defined for this high-energy case. As detailed in Table \ref{tab:detailed_results}, the GA's solution requires a total $\Delta V$ of 3887.15 m/s. This represents a substantial saving of 213.47 m/s compared to the Hohmann transfer's cost of 4100.62 m/s. This result empirically demonstrates the superiority of the Bi-elliptic strategy for high-energy transfers.

\begin{figure}[!t]
    \centering
    \includegraphics[width=0.8\textwidth]{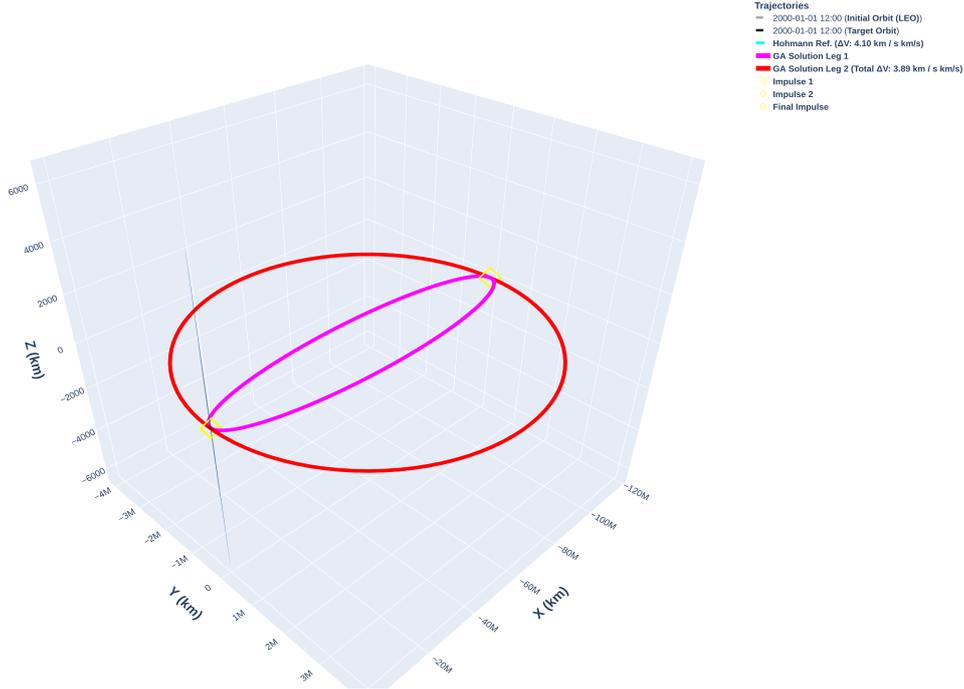}
    \caption{Wide view of the optimal trajectory for the LEO-to-Far-Orbit scenario. The GA's Bi-elliptic path (magenta and red segments) is shown to scale, highlighting its immense apoapsis. The Hohmann transfer (cyan) is contained entirely within the central region.}
    \label{fig:far_orbit_wide}
\end{figure}

\begin{figure}[!t]
    \centering
    \includegraphics[width=0.8\textwidth]{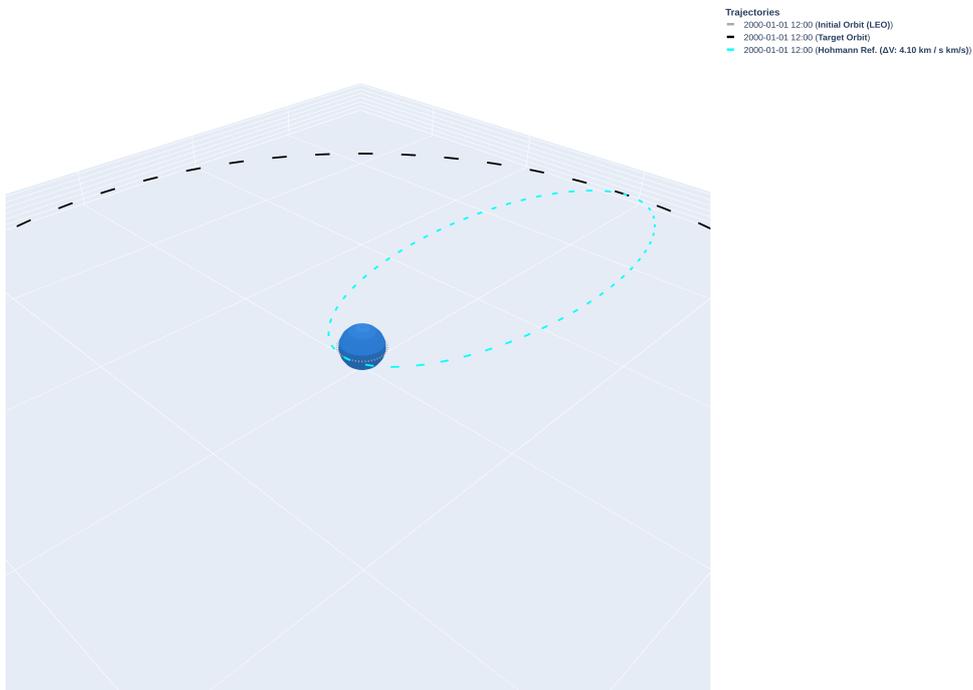}
    \caption{Zoomed-in view for the LEO-to-Far-Orbit scenario. This plot clearly shows the initial LEO, the final target orbit, and the classical Hohmann transfer path (cyan, dashed) for comparison.}
    \label{fig:far_orbit_zoom}
\end{figure}

This fuel efficiency, however, comes at a significant trade-off. The GA's optimal path requires over 52,000 days (more than 140 years) to complete, whereas the Hohmann transfer takes just over one day. This highlights a critical trade-off between propellant mass and mission duration that our framework effectively quantifies. The immense scale of this trajectory is visualized in Figures \ref{fig:far_orbit_wide} and \ref{fig:far_orbit_zoom}.

\begin{table}[H]
    \centering
    \caption{Detailed Performance Metrics for GA-Optimized and Hohmann Transfers. All data is now correctly aligned with its respective scenario, and numerical precision is consistent.}
    \label{tab:detailed_results}
    \begin{tabular}{l c c}
        \toprule
        & \textbf{Scenario 1:} & \textbf{Scenario 2:} \\
        \textbf{Metric} & \textbf{LEO to GEO} & \textbf{LEO to Far Orbit} \\
        \cmidrule(lr){2-2} \cmidrule(lr){3-3}
        \multicolumn{3}{l}{\textbf{GA-Optimized Solution}} \\
        \quad Intermediate Radius ($r_b$) [km] & 42,164.00 & 118,190,508.83 \\
        \quad Total $\Delta V$ [m/s] & 3853.96 & 3887.15 \\
        \quad Total Time [days] & 0.22 & 52,374.31 \\
        \midrule
        \multicolumn{3}{l}{\textbf{Hohmann Transfer (for comparison)}} \\
        \quad Total $\Delta V$ [m/s] & 3853.96 & 4100.62 \\
        \quad Total Time [days] & 0.22 & 1.09 \\
        \bottomrule
    \end{tabular}
\end{table}

The key findings from our experiments are concisely summarized in Table \ref{tab:summary}. This table highlights the adaptive capability of the GA-based framework, showing it correctly identifies the optimal strategy for each energy regime.

\begin{table}[H]
    \centering
    \caption{Summary of Optimal Strategy and $\Delta V$ Savings.}
    \label{tab:summary}
    \begin{tabular}{lcc}
        \toprule
        \textbf{Scenario} & \textbf{Optimal Strategy Found by GA} & \textbf{$\Delta V$ Savings (m/s)} \\
        \midrule
        LEO to GEO ($r_f/r_i \approx 6.2$) & Hohmann Transfer & 0.00 \\
        LEO to Far Orbit ($r_f/r_i = 20$) & Bi-elliptic Transfer & 213.47 \\
        \bottomrule
    \end{tabular}
\end{table}

\subsection{Discussion}

The experimental results demonstrate the efficacy and adaptability of our GA-based optimization framework in autonomously identifying fuel-optimal orbital transfer strategies.

\paragraph{Validation of Classical Solutions.} In the LEO-to-GEO scenario ($r_f/r_i \approx 6.2$), the GA consistently converged to a total $\Delta V$ of 3853.96 m/s, precisely matching the analytical Hohmann transfer (see Table \ref{tab:detailed_results}). This result validates the accuracy of our framework, as the GA rediscovered the theoretically optimal solution without prior knowledge. The convergence to $r_b/r_f \approx 1.00$ confirms the algorithm correctly identified the degenerate case of the Bi-elliptic transfer where the third impulse is zero.

\paragraph{Superiority in High-Energy Regimes.} For the LEO-to-Far-Orbit scenario ($r_f/r_i = 20$), the GA identified a Bi-elliptic transfer that achieved a savings of 213.47 m/s compared to the Hohmann transfer (Table \ref{tab:summary}). This aligns with classical theory, which predicts Bi-elliptic transfers become more fuel-efficient when $r_f/r_i > 11.94$ \citep{Sternfeld1934}. The discovery of this solution underscores the GA’s ability to explore extreme regions of the search space, a capability enabled by the adaptive bounds discussed in the methodology.

\paragraph{Trade-Off Between Fuel and Time.} The substantial $\Delta V$ savings in the high-energy scenario are accompanied by an impractical transfer time of over 140 years. While this renders the solution infeasible for most time-critical missions, it may be relevant for niche applications where flight time is a secondary constraint, such as for robotic interstellar precursor probes, long-term cycler concepts, or certain orbital debris disposal strategies. This trade-off suggests that future iterations could incorporate multi-objective optimization to identify a Pareto front of solutions balancing $\Delta V$ and transfer time.

\paragraph{Practical Significance and Accessibility.} The framework’s ability to rediscover classical solutions and identify superior strategies highlights its potential for practical applications. By leveraging open-source tools, the methodology lowers the barrier to entry for astrodynamics research, enabling high-fidelity trajectory optimization with modest computational resources and enhancing its utility for rapid prototyping and iterative mission design.

\paragraph{Limitations and Future Directions.} While the framework excels in the idealized scenarios presented, its extension to more complex problems offers a clear path for future work. The assumption of instantaneous impulses, for example, could be replaced by finite burns to model low-thrust trajectories. This could be achieved by parameterizing the thrust profile (e.g., as a series of polynomial splines) and including these parameters in the GA's chromosome. Similarly, to handle non-coplanar transfers, the chromosome could be expanded to include variables for an out-of-plane burn (e.g., its magnitude and location), with the fitness function modified to account for the additional inclination change. Such extensions would make the framework more versatile for operational mission planning.

In summary, the results validate the proposed framework as a powerful and accessible tool for trajectory optimization, capable of autonomously discovering optimal strategies while quantifying critical trade-offs.

\section{Conclusion}

In this work, we have successfully developed and demonstrated a computational framework that integrates a Genetic Algorithm with the Poliastro simulation library to autonomously find fuel-optimal, three-impulse orbital transfers. Our experiments have shown that this approach is not only effective but also highly adaptive to the specific energy requirements of a given mission.

The key findings are twofold. First, for the low-energy LEO-to-GEO transfer, our GA-based optimizer precisely rediscovered the classical Hohmann transfer as the optimal solution, thereby validating the accuracy and reliability of our methodology. Second, when presented with a high-energy transfer to a distant orbit, the same framework autonomously discovered a superior Bi-elliptic trajectory, yielding a significant fuel saving of 213.47 m/s over the Hohmann alternative. This result empirically confirms established astrodynamical theory while also quantifying the extreme trade-off between propellant efficiency and flight duration, which increased from one day to over a century.

The primary strength of this work lies in its demonstration of a flexible and accessible optimization toolchain. By combining a robust heuristic search algorithm with a readily available, open-source physics engine, we have created a system capable of solving non-trivial astrodynamics problems without requiring complex analytical derivations. A clear path for future work involves extending this framework to higher-dimensional, more realistic problems. This includes optimizing non-coplanar transfers and finite-burn, low-thrust trajectories by expanding the problem's genetic encoding, as well as formally solving for the trade-off between $\Delta V$ and transfer time via multi-objective optimization. This study serves as a strong testament to the growing potential of computational intelligence in modern space mission design.



\bibliographystyle{plainnat} 

\end{document}